# BLOFF: A Blockchain based Forensic Model in IoT

**Promise Agbedanu**

**Donghua University:** Shanghai, CN, email: ricardopromise@gmail.com
https://orcid.org/0000-0003-2522-891X

**Anca Delia Jurcut**

University College Dublin, Ireland, email: anca.jurcut@ucd.ie
https://orcid.org/0000-0002-2705-1823

**ABSTRACT**

*In this era of explosive growth in technology, the internet of things (IoT) has become the game changer when we consider technologies like smart homes and cities, smart energy, security and surveillance, and healthcare. The numerous benefits provided by IoT have become attractive technologies for users and cybercriminals. Cybercriminals of today have the tools and the technology to deploy millions of sophisticated attacks. These attacks need to be investigated; this is where digital forensics comes into play. However, it is not easy to conduct a forensic investigation in IoT systems because of the heterogeneous nature of the IoT environment. Additionally, forensic investigators mostly rely on evidence from service providers, a situation that can lead to evidence contamination. To solve this problem, the authors proposed a blockchain-based IoT forensic model that prevents the admissibility of tampered logs into evidence.*

## 1 Introduction

In this era of explosive growth in technology, the Internet of Things (IoT) has become the game changer when we consider technologies like smart homes and cities; smart energy, security and surveillance and healthcare. In a report, Statista predicted that the number of IoT device will reach 75 billion in 2025 (Statista, 2019). The integration of real-world objects with the internet does not only bring numerous advantages but also bring cybersecurity threats to our life, through our interaction with these devices (A. Jurcut et al., 2020). Like any computing technology, IoT is threatened by security issues. Many researchers and device manufacturers are exploring various techniques to ensure the security of IoT devices as well as protect the data generated by these devices. However, according to (Atlam et al., 2017), it is difficult to secure data produce in IoT environments because of the heterogeneity and dynamic features deployed in these devices. It is therefore not surprising that the security of IoT, ranging from the physical security of the devices through to the security of their architecture has become an important area of research for a lot of researchers (A. D. Jurcut et al., 2020).

Currently, several works are being done to ensure data confidentiality, access control, authentication, privacy and trust in IoT environments (Borhani et al., 2020; Braeken et al., 2019; A. Jurcut et al., 2009, 2012; A. D. Jurcut, 2018; A. D. Jurcut et al., 2014; Kumar et al., 2019; Xu et al., 2019). Although a lot of success has been made in the security of IoT using the parameters mentioned above, attackers still find ways to exploit the vulnerabilities that exist in IoT systems (A. D. Jurcut et al., 2020). These billions of IoT devices contain sensitive data, an attribute that makes them attractive to cyber-attacks. The number and the cost of cyber-attacks have been increasing over the years. According to a report by (Morgan, 2017), the damages caused by cybercrimes will cost a whopping of 6 trillion dollars by 2021. These attacks need to be investigated; this is where digital forensics comes into play.

Digital forensics helps in acquiring legal evidence that can be used to more about these attacks, prevent future attacks and most importantly prosecute the perpetrators of these crimes. However, it is not easy to conduct a forensic investigation in IoT systems. According to (Perumal et al., 2015), the heterogeneity and dynamic nature of IoT systems make it practically difficult to use the same frameworks used in traditional digital forensic in IoT environments. It is therefore expedient to develop frameworks that can be used in IoT environments considering their dynamic and heterogeneity nature.

In this chapter, we discuss a blockchain-based model that ensures the verifiability of logs produced in IoT environments. The main idea of this model is to ensure the credibility and authenticity of logs produced by IoT devices during forensic investigations. Our model uses the decentralized approach and the immutability property of blockchain to ensure that logs and other pieces of evidence produced in IoT environments can be verified by forensic stakeholders. Our proposed model prevents Cloud Service Providers (CSPs) or Law Enforcement Agencies (LEAs) from tendering in false evidence during forensic investigations or court proceedings. The proposed model brings the court, LEAs, CSPs and other stakeholders under one umbrella where each stakeholder can verify the authenticity of the evidence presented by any of them. The model also ensures that pieces of evidence are not tampered with during the chain of custody. We start this chapter by providing an introduction to digital forensics in IoT and blockchain. We then follow by describing our proposed model. The benefits of our model are discussed and then we present our conclusion.

## 2. Background

In this section, we present a background of digital forensics, IoT forensics and blockchain.

### 2.1 Digital Forensics

Several definitions have been given for digital forensics. (Bellegarde et al., 2010) defined computer forensics as "the preservation, identification, extraction, interpretation, and documentation of computer evidence to include the rules of evidence, legal process, the integrity of evidence, factual reporting of the information and providing expert opinion in a court of law or other legal and/or administrative proceedings as to what was found." The concept of digital forensics and computer forensics are closely related with the latter being the subset of the former. The most widely accepted definition of digital forensics is the one proposed during the first digital forensic research workshop and was stated by authors in (James et al., 2015) as "The use of scientifically derived and proven methods toward the preservation, collection, validation, identification, analysis, interpretation, documentation and presentation of digital evidence derived from digital sources to facilitate or further the reconstruction of events found to be criminal, or helping to anticipate unauthorized actions shown to be disruptive to planned operations." Digital forensics involves the application of scientific methods in investigating cyber-crimes (Carrier, 2003). According to (Sumalatha & Batsa, 2016), the methodologies employed in digital forensics to handle electronic evidence has its stems springing out from forensic science.

**2.1.1 Digital Forensic Process**

There are various stages that digital forensic artifacts undergo before finally being presented in a court for prosecution purposes. (Daniel & Daniel, 2012) stated that the digital forensic is made up of four processes. These processes are identification, collection, organization, and presentation. Similarly, (Zawoad et al., 2015) enumerated the stages of digital forensics as identification, preservation, analysis, and presentation. According to (Hemdan & Manjaiah, 2018), digital forensics involves the application of scientific processes in identifying, collecting, organizing and presenting evidence. Ken Zatyko, the a former director of the US Defense Computer Forensics Laboratory outlined an eight-step process that makes digital forensic a scientific process (Zawoad et al., 2015). These eight steps include (Zawoad et al., 2015): obtaining the search authority, documenting the chain of custody, imaging, and hashing of evidence, validating tools used in the forensic process, analysing evidence, repeating and reproducing to ensure quality assurance, report by documenting the procedures used in the forensic process and then finally, present an expert witness in a court of law.

**2.1.2 Evidence Identification**

This is the first stage of the forensic process. This stage involves two steps. The first is the identification of the incident and the second is the identification of the evidence. There must be a direct correlation between the incident and the evidence being identified (Daniel & Daniel, 2012).

**2.1.3 Evidence Collection**

This stage involves the extraction of evidence from different media. The extraction methods may include imaging the original copy of the evidence. This stage also involves preserving the integrity of the evidence (Daniel & Daniel, 2012).

**2.1.4 Organization**

This stage has two main steps including evidence of examination and evidence analysis. Some researchers separate these two steps distinctively. During the evidence examination, the investigator performs a thorough inspection of the data being used as evidence (Daniel & Daniel, 2012). This inspection may involve the use of different forensic tools. These tools are used for extracting and filtering data that is of interest to the investigator and relevant to the investigation process (Zawoad & Hasan, 2015). The analysis phase involves reconstructing events by analysing the data collected. The rationale behind the evidence analysis is to discover any evidential material which will aid the technical as well as the legal perspective of the case (Ademu et al., 2011).

### 2.1.5 Evidence Presentation

The evidence, after the identification, collection and organization; needs to be presented to a court of law. This stage includes the investigator preparing an organized report to state the findings he or she made during the investigation process (Daniel & Daniel, 2012). These findings are then presented to a court of law with the investigator serving as an expert witness if there is a need to testify (Ieong, 2006).

### 2.2 IoT Forensics

Unlike the traditional digital forensics, IoT forensics is a new and unexplored area by both industry and academia. Although the purpose of both digital and IoT forensics, is to extract digital information using a scientific approach; the scope available when it comes to information extraction is wider in IoT forensics. According to (Atlam et al., 2017; Raghavan, 2013), IoT forensics is made up of the cloud, network and the device level forensics. Similarly, (Stoyanova et al., 2020) defined IoT forensics as an aspect of digital forensics with the identification, collection, organization, and presentation of evidence happening within the IoT ecosystem. They also broke IoT forensics into Cloud, Network, and Device-level forensics.

### 2.3 Blockchain

Blockchain, also known as distributed ledger technology is a decentralized and distributed ledger that contains chains of blocks of various transactions joined together by cryptographic hashes. According to (Gaur et al., 2018), the blockchain is "an immutable ledger for recording transactions, maintained within a distributed network of mutually untrusted peers." Any transaction coming from a node is validated by other participating nodes in the blockchain network. After the validation, the set of transactions is added to the block by special nodes called miners as in the case of bitcoin. A miner is a node with sufficient computational power to solve a cryptographic puzzle. Blockchain uses a peer-to-peer (P2P) network. This architecture makes it possible for each node to communicate with a set of neighbour nodes, then each of these nodes also communicate with their neighbour nodes and the communication goes on and on. The blockchain is designed in such a way that any node can join and leave the network at will. Certain key elements are needed in the design and implementation of blockchain technology. These elements are:

1. **Timestamping:** This means that the problem of double-spending in the case of cryptocurrency applications like bitcoin is avoided. Timestamping is achieved by collecting pending transactions into the block and then calculating the hash of the block. This can prove that a transaction existed at the time of creating the block since it is hashed onto the block.

2. **Consensus:** Because new blocks are created and broadcasted by mining nodes, all nodes need to agree on a single version of the block. A distributed consensus helps to decide on which block out of the several variants generated by different nodes would be added to the blockchain.

3. **Data security and integrity:** This property or attribute prevents a malicious node from creating a fake transaction since each transaction is signed by a node or a user using their private key. Similarly, (Gaur et al., 2018) also identifies four blocks within a blockchain framework. These blocks are: the shared ledger, cryptography, consensus, and the smart contracts.

Mining also called proof-of-work (PoW) is used to achieve consensus and to ensure data security and integrity. Mining is done based on the sequence of transactions; a situation can only be changed by redoing the proof-of-work. Mining introduces the difficulty in block generation. There are other methods available ensuring data security and integrity. Some of these methods are proof-of-authority (PoA), proof-of-existence (PoE) and proof-of-concept (PoC). A blockchain can either be permissionless or permissioned. A permission-less blockchain also known as public blockchain allows any node to join and leave at any point in time, whereas private or permissioned blockchain allows nodes to be authenticated before joining the network.

## 3 Related Work

In this section, we explore some works that are closely related to ours. Blockchain has been widely used in the area of IoT security. However, the concept is still in its exploratory stage when it comes to IoT forensics. (Meffert et al., 2017) proposed a framework that helps evidence acquisition in IoT forensic. Using a centralized approach, the framework is deployed in three nodes namely; controller to IoT device, controller to cloud and controller to controller. Although, the proof of concept used in this work showed that the proposed framework can pull forensic data from IoT devices. However, the centralized nature of their approach makes it difficult to authenticate the evidence captured by the framework. (Li et al., 2019) also used a digital witness approach that allows people to shared logs from IoT devices with guaranteed privacy. The authors used the Privacy-aware IoT Forensics (PRoFIT) model proposed in their earlier work to deploy their digital witness model. The method proposed in their work is to help collect digital evidence in IoT environments as well as ensuring that the privacy of the evidence collected is maintained. The proposed method supports 11 privacy principles captured (Nieto et al., 2018) in their PRoFIT methodology. In their work, (Nieto et al., 2017) proposed a distributed logging scheme for IoT forensics. In this work, the authors used a Modified Information Dispersal Algorithm (MIDA) that ensures the availability of logs generated in IoT environments. The logs are aggregated, compressed, authenticated and dispersed. The distributed approach used in this work only focuses on how logs are stored but not how they are verified. Leveraging the immutability property of blockchain technology, (Noura et al., 2020) proposed an IoT forensics framework that uses a permissioned-based blockchain. This framework enhances the integrity, authenticity and non-repudiation of pieces of evidence.

## 4 Proposed Model

In this section, we present our blockchain-based forensic model for IoT (BLOF). In IoT environments, there are three layers. These are the cloud, network and the device layer. Our model leverages the decentralized property of blockchain to ensure that the logs produced in IoT environments are stored on the network and are available for the verification by any of the participating nodes in the network. There are several artefacts to be considered when conducting a forensic investigation. However, our model focuses only on system and event logs. The entities in our model are the Cloud Service Providers (CSPs), Network Devices and the IoT devices. The entities serve as the blockchain nodes in the network. New nodes are added to the network through a key generation process. The public key of a node is appended to a transaction before it is written onto the block. New nodes generate a pair of keys. The CSPs act as the miners in the network. Their computational capacity of CSPs makes them ideal candidates for mining. Our model is made up of a Blockchain Centre (BC), Log Processing Centre (LPC) and User Centre (UC). In the preceding subsections, we discuss each of these components into details. The proposed model is shown in Figure 1.

### 4.1 Blockchain Centre

The blockchain centre is made up of a distributed ledger where each log is written onto a block after processing. The distributed ledger is made up of all blocks that have been committed to the network. Each block contains a transactional value of hashed values computed from logs. The logs are extracted from the various entities, hashed and written onto the blockchain network as transactions. The nodes in the BC are made up of forensic stakeholders, network and IoT devices and cloud service providers. The blocks are proposed after a consensus has been reached by the nodes.

### 4.2 Log Processing Centre

This component of our model handles the log processing. The LPC is an Application Programming Interface (API) that sits between entities and the blockchain network. The LPC extracts the logs from the IoT devices, network devices and the cloud layers. The extracted logs are hashed using a SHA-256 hash function and the hashed values are written onto the block as a transaction. We chose to hash the logs because these logs may contain sensitive information. Therefore, it is not advisable to store the logs as plaintext. Secondly, hashing the logs reduces its size which eventually reduces the time needed to process the logs.

$$Transaction = hash\,(log)$$

### 4.3 User Centre

The user centre is made up of the courts and the forensic investigators. This component of the model makes it possible for forensic investigators to verify the authenticity of logs presented to them by service providers. Additionally, forensic investigators can verify the authenticity of logs even as they are passed on from one investigator to another during the chain of custody. Furthermore, the court can also verify the authenticity of logs presented by prosecutors and decide if such a piece of evidence (log) must be admitted or not. This prevents investigators from tampering with logs to either incriminate innocent people or to exonerate criminals.

## 5 Verification of Logs

Unlike the traditional digital forensics, investigators solely depend on the CSPs, network and the IoT devices for evidence when it comes to the IoT forensics. This dependence on CSPs may lead to compromising pieces of evidence. Our proposed model does not only ensure the integrity of logs through the use of a decentralized ledger but also allows logs tendered in as evidence to be verified by various stakeholders in the forensic process.

In our proposed model, the forensic investigator still falls on the CSP and IoT devices for the evidence. However, the evidence which in this case is the various logs generated by the cloud instances, network and the IoT devices are hashed after a forensic investigator receives such evidence from a CSP. The hashed value is then compared to the hashes stored as transaction values on the blockchain network. The investigator then searches for the hashed value on the blockchain network. If the hash value exists on the blockchain then the log is accepted by the investigator and forwarded to the court as credible evidence. On the other hand, if the hash value does not exist on the blockchain network then the log is rejected. When the court receives a log from a forensic investigator, the court can determine the credibility of the log by similarly hashing the log and comparing the hashed value to the hash values on the blockchain network. If the value exists, the evidence is accepted by the court. Otherwise, it is rejected. The verification process of our propose model is shown in Figure 2.

For example, Bob who is a forensic investigator receives logs from a CSP intending to present the evidence to a court to prosecute an attacker called Elvan. However, Bob is not sure if the logs provided by the CSP has been tampered with or not. Since the hash of all logs is processed and stored in our blockchain-based model; Bob can verify the authenticity of the logs. Bob first runs the log through a SHA-256 hash function and gets a value (x). He then searches for the transaction (x) on the blockchain network. Bob is one of the participating nodes on the network. If he finds the value (x) as a valid transaction on the network; then the log is genuine and he can then proceed to court and present the log as a piece of evidence. However, if the value (x) is not a valid transaction then Bob must discard that log and rather investigate who made changes to the log. After Bob is presenting the log to the court, the court can also verify the authenticity of the log. The assumption here is that Bob might try to implicate Elvan as the perpetrator of a crime he is innocent of. Similarly, the court runs the log through a SHA-256 hash function and then search for the exact value on the blockchain network. If the transaction value exists, then the court admits the log as evidence and proceeds with hearing the case. Otherwise, the court rejects the evidence and dismisses the case.

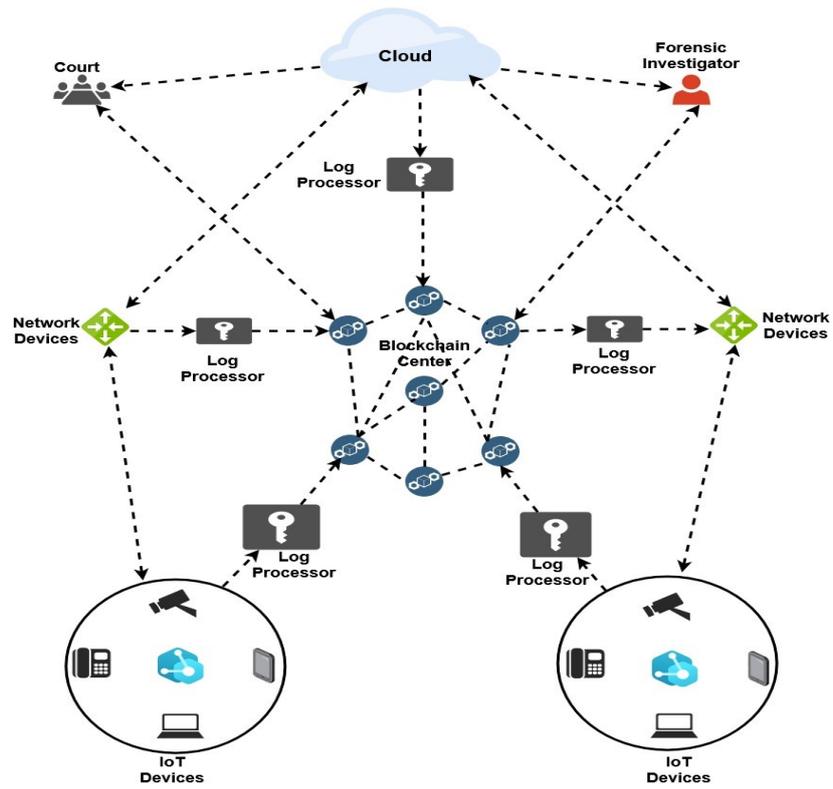

**Figure 1.** Our proposed blockchain-based IoT forensic model

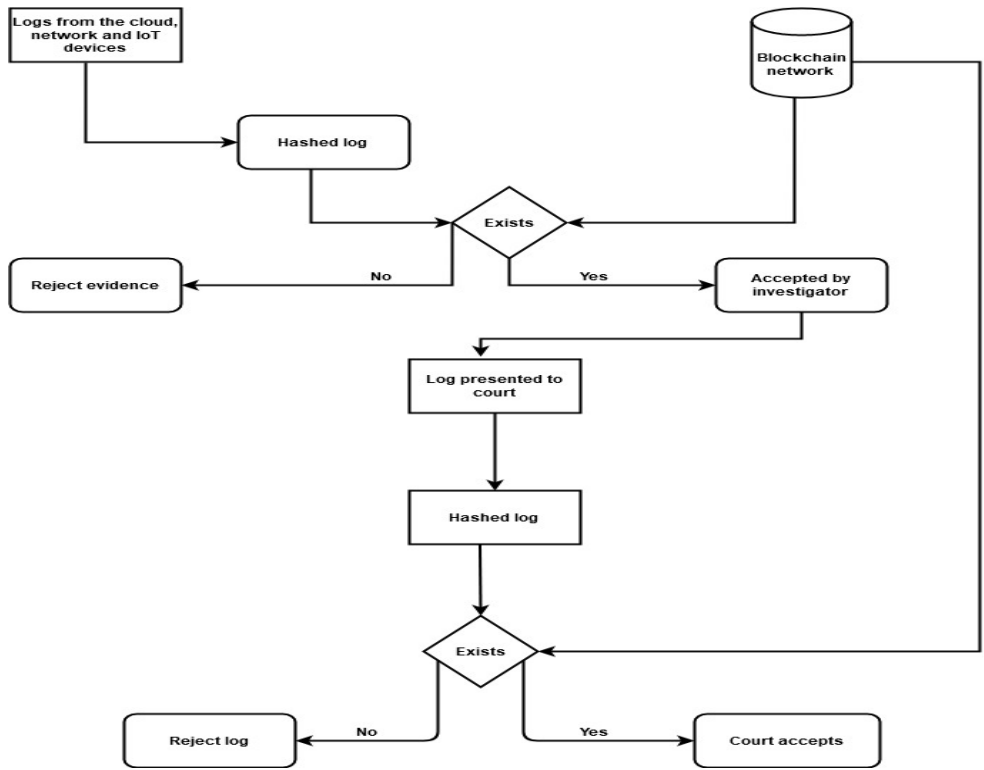

**Figure 2.** Verification process of our proposed model

## 6 Discussion

In this section we discuss the advantages of our proposed model, the possible impact it may have on the performance of IoT forensics and then we compare our model to the current existing ones.

Firstly, our proposed model is based on blockchain, making the model a fully decentralized one as compared to a centralized approach discussed in (Meffert et al., 2017). The decentralized characteristic of our model ensures that logs can be verified to determine their authenticity or otherwise. It also prevents service providers and forensic investigators from tampering with logs without being detected. Our model uses the immutability property of blockchain as a leverage to ensure the integrity of logs produced in the IoT environment.

Further, our model offers the advantage of verifiability. This advantage gives the possibility for the forensic stakeholders to verify the authenticity of the logs produced in IoT environments - an advantage that is either not available or not fully explored in the existing related work (Nieto et al., 2018; Nieto et al., 2017; Nieto et al., 2017; Le et al., 2018).

We compare our model with the current work that is closely related to ours, using the following parameters: blockchain-based, verifiability, decentralized, evidence integrity and privacy. In the instance where a model satisfies a parameter, we mark it as a Yes (Y). If it does not satisfy that parameter then we mark it as a No (N). The comparison is shown in Table 1.

**Table 1.** Comparing our model to other existing models

| Metric | (Nieto et al., 2018) | (Nieto et al., 2017) | (Noura et al., 2020) | (Le et al., 2018) | Our Model |
|---|---|---|---|---|---|
| Blockchain-based | N | N | N | Y | Y |
| Verifiability | N | N | N | N | Y |
| Decentralized | N | N | N | Y | Y |
| Evidence integrity | Y | N | Y | Y | Y |
| Privacy | Y | Y | Y | Y | Y |

We acknowledge the fact that integrating our model into an IoT environment will come with some additional cost in terms of resource usage and computation. In our subsequent work, we intend to evaluate the performance of our model by setting up a testbed to validate the effectiveness of the model.

## 7 Conclusion

Today, IoT has become an integral part of human life. It is integrated into domains such as healthcare, automobile, agriculture, manufacturing industry and household. The world is going to see greater growth in IoT technology with the introduction of 5G network. However, just like any computing technology security of this technology is a concern. With an exponential increase in the number of cyber-attacks, it is important that such crimes are investigated and that the perpetrators are brought to justice. Due to the heterogeneous nature of the IoT environment coupled with the integration of the cloud and the network layer, makes the forensic investigations in an IoT environment a very challenging task. Further, it is extremely difficult for the stakeholders to determine the authenticity of the evidence they deal with, since in most cases they have to depend on service providers for these pieces of evidence. To ensure that logs presented to forensic investigators are authentic and tamper-free, we proposed a blockchain forensic model that uses a decentralized approach to keep the hashed values of logs produced in IoT environment as transactional records. The proposed model allows any forensic stakeholder to verify the authenticity of the logs they are working with. This model ensures that innocent people are not framed up and culprits are exonerated by interested parties during a forensic investigation. For future work, we seek to perform an experimental validation to ascertain the computational impact our model will have in an IoT environment. Additionally, exploring ways of storing logs within our model is another area that is of interest to us.

bibliographyhttps://books.google.co.za/books?id=wKdhDwAAQBAJ

Hemdan, E. E., & Manjaiah, D. H. (2018). *CFIM : Toward Building New Cloud Forensics Investigation Model*. 545–554.

Ieong, R. S. C. (2006). FORZA–Digital forensics investigation framework that incorporate legal issues. *Digital Investigation*, *3*, 29–36.

James, J. I., Shosha, A. F., & Gladyshev, P. (2015). Digital Forensic Investigation and Cloud Computing. In *Cloud Technology* (pp. 1231–1271). IGI Global. https://doi.org/10.4018/978-1-4666-6539-2.ch057

Jurcut, A., Coffey, T., & Dojen, R. (2012). Symmetry in Security Protocol Cryptographic Messages--A Serious Weakness Exploitable by Parallel Session Attacks. *2012 Seventh International Conference on Availability, Reliability and Security*, 410–416.

Jurcut, A., Coffey, T., Dojen, R., & Gyorodi, R. (2009). Security Protocol Design: A Case Study Using Key Distribution Protocols. *Journal of Computer Science & Control Systems*, *2*(2).

Jurcut, A. D. (2018). Automated logic-based technique for formal verification of security protocols. *Journal of Advances in Computer Network*, *6*, 77–85.

Jurcut, A. D., Coffey, T., & Dojen, R. (2014). Design requirements to counter parallel session attacks in security protocols. *2014 Twelfth Annual International Conference on Privacy, Security and Trust*, 298–305.

Jurcut, A. D., Ranaweera, P., & Xu, L. (2020). Introduction to IoT Security. *IoT Security: Advances in Authentication*, 27–64.

Jurcut, A., Niculcea, T., Ranaweera, P., & LeKhac, A. (2020). Security considerations for Internet of Things: A survey. *ArXiv Preprint ArXiv:2006.10591*.

Kumar, T., Braeken, A., Jurcut, A. D., Liyanage, M., & Ylianttila, M. (2019). AGE: authentication in gadget-free healthcare environments. *Information Technology and Management*, 1–20.

Le, D.-P., Meng, H., Su, L., Yeo, S. L., & Thing, V. (2018). Biff: A blockchain-based iot forensics framework with identity privacy. *TENCON 2018-2018 IEEE Region 10 Conference*, 2372–2377.

Li, S., Choo, K.-K. R., Sun, Q., Buchanan, W. J., & Cao, J. (2019). IoT forensics: Amazon echo as a use case. *IEEE Internet of Things Journal*, *6*(4), 6487–6497.

Meffert, C., Clark, D., Baggili, I., & Breitinger, F. (2017). Forensic State Acquisition from Internet of Things (FSAIoT): A General Framework and Practical Approach for IoT Forensics Through IoT Device State Acquisition. *Proceedings of the 12th International Conference on Availability, Reliability and Security*, 56:1--56:11. https://doi.org/10.1145/3098954.3104053

Morgan, S. (2017). *Cybercrime Report, 2017*.

Nieto, A., Rios, R., & Lopez, J. (2018). IoT-forensics meets privacy: towards cooperative11